# Flexural Plate Wave Devices

# Fabricated from Silicon Carbide Membrane

*

Ndeye Fama Diagne[1,2], James Lindesay[1,2], and Michael Spencer[2]
[1]Computational Physics Laboratory
[2]Material Science Research Center of Excellence
Howard University, Washington, D.C.

## ABSTRACT

Flexural Plate Wave (FPW) devices fabricated from Silicon Carbide (SiC) membranes are presented here which exhibit electrical and mechanical characteristics in its transfer functions that makes it very useful as a low voltage probe device capable of functioning in small areas that are commonly inaccessible to ordinary devices. The low input impedance characteristic of this current driven device makes it possible for it to operate at very low voltages, thereby reducing the hazards for flammable or explosive areas to be probed. The Flexural Plate Wave (FPW) devices are of a family of gravimetric type sensors that permit direct measurements of the mass of the vibrating element. The primary objective was to study the suitability of Silicon Carbide (SiC) membranes as a replacement of Silicon Nitride (SiN) membrane in flexural plate wave devices developed by Sandia National Laboratories.

Fabrication of the Flexural Plate Wave devices involves the overlaying a silicon wafer with membranes of 3C-SiC thin film upon which conducting meander lines are placed. The input excitation energy is in the form of an input current. The lines of current along the direction of the conducting Meander Lines Transducer (MLTs) and the applied perpendicular external magnetic field set up a mechanical wave perpendicular to both, exciting the membrane by means of a Lorentz force, which in turn sets up flexural waves that propagate along the thin membrane. The physical dimensions, the mass density, the tension in the membrane and the meander spacing are physical characteristics that determine resonance frequency of the Flexural Plate Wave (FPW) device. Equivalent circuit models characterizing the reflection response $S_{11}$ (amplitude and phase) for a one-port Flexural PlateWave device and the transmission response $S_{21}$ of a two-port device are used for the development of the equivalent mechanical characteristics.

# CHAPTER I

## Introduction

### 1.1    Introduction

The flexural plate wave (FPW) device is a micro-machined thin film device capable of being configured into and used as an acoustic sensor.   The key sensing element in the Flexural Plate Wave sensor is a thin "plate" along which flexural plate waves propagate[1].  The thickness of the plate is much smaller than the wavelength of the ultrasonic waves propagating in it, and therefore the wave propagates at a velocity that is lower than the speed of sound.

The plate is so thin that its elastic response can be considerably influenced by in-plane tension that develops during the initial fabrication process.  The mechanical and material properties of the membrane allow external forces and stress to alter the behavior of the device. Plate waves, unlike surface acoustic waves,  have a phase velocity that depends on the thickness of the propagating medium.

One of the unique features of these Flexural Plate Wave devices is that silicon carbide is a wide bandgap semiconductor material.  Also silicon carbide has a very high thermal conductivity, which makes it attractive for high temperature devices.

The Flexural Plate Wave device is driven by an external current to create a Lorentz force, which in turn induces a voltage on the second (output port) because of the coupling of mechanical flexural waves to electrical waves.  Different transduction schemes allow us to excite particular modes and to measure their properties directly or indirectly.  Transduction can be divided into one-port and two-port schemes, each with passive and active (oscillator) implementations.  Two-port configurations can be used to measure wave velocities and attenuation in several ways.  For example, phase velocity shifts can be inferred from the phase shift between input and output transducers.   Phase can be measured passively at a constant operating frequency or actively using a feedback or phase shift oscillator.   The group velocity can be calculated from the measured  group delay time between input and output transducers, or with phase modulation techniques.  The dissipation or attenuation can be found from changes of insertion loss (the ratio of output to input voltage or current).

One-port schemes can yield similar information.  For example, the impedance of the transducer can be measured as a function of frequency to find resonant frequencies of the device and information about energy dissipation or wave attenuation.  It is also common to track the natural frequencies actively using an oscillator.

Figure 1.1 is a classification of ultrasonic measurement methods.  As the characteristics of the path are altered, so are wave properties, such as amplitude, frequency, phase or group velocities, phase, etc...We measure changes of these properties to get information about the measurand.

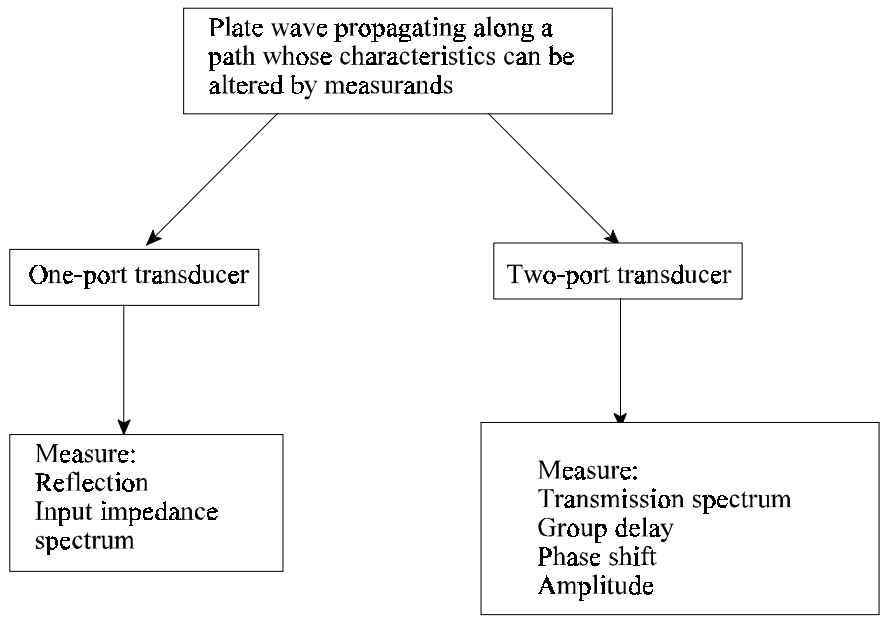

Fig. 1.1  Classification of Ultrasonic Measurement Methods

## 1.2    Objectives

The objective of this research was to study the suitability of Silicon Carbide films (SiC) as a replacement of Silicon Nitride films (SiN) in the magnetically-excited flexural plate wave device developed by Sandia National Laboratories, by designing, fabricating and characterizing Flexural Plate Wave devices.

The motivation for this research was to take advantage of the physical properties of Silicon Carbide. These properties include superior membrane characteristics such as hardness, stiffness, high temperature elasticity, high thermal conductivity, very low oxidation and corrosion rates, and extreme chemical inertness. At the time of this writing, there was a very little or no published data on some of the needed parameters (for example resonance frequency) and there were no publications on Silicon Carbide (SiC) Flexural Plate Wave devices.

Chapter 1 provides a brief literature review and a summary of the Flexural Plate wave devices. Chapter 2 gives a discussion on the flexural plate wave device application. Chapter 3 gives a discussion of the theory of plate wave devices, the differential wave equation, the linear response of the FPW device, and the principal of operation of the Flexural Plate wave devices. This chapter discusses also the FPW device design, mask layout and Meander Line Transducers(MLT). Chapter 4 discusses the properties of Silicon Nitride (SiN) and Silicon Carbide (SiC) films. Chapter 5 describes the FPW device fabrication process. Chapter 6 gives a discussion on the techniques used to characterize the FPWs devices. Chapter 7 presents the results obtained from this study and provides some discussion of these results. Chapter 8 provides a summary and suggestions of future research.

## 1.3     Literature Review

The elastic wave deices that have been studied most extensively are surface acoustic wave (SAW) delay lines, bulk acoustic wave (BAW) resonators, thickness-shear mode (TSM) resonators, and  flexural plate wave (FPW) devices.  The majority of applications that have been studied involved sensing.  There are large overlaps between the specific applications, but generally the ultrasonic devices have been used more as gravimetric or mass sensors, with applications to chemical vapor sensing and film thickness monitoring, while resonant structures have been used more  for detecting force, acceleration and pressure.  Surface waves have been extensively used for sensing chemical vapor,[45-58] and for measuring force, acceleration, pressure[59-64] as well as the properties of viscoelastic materials[65-66] and liquids [67].  The use of SAW devices as gravimetric bio-sensors in liquids for detecting antibodies [68-71] has been only partially successful because of the inherently high attenuation caused by compressional wave radiation.

The first established sensing application for BAW resonators was in monitoring film thickness during deposition processes [72-73].  Since then BAW devices have been used for vapor sensing, studying chemical and electrochemical processes and interactions such as corrosion and absorption  in thin films[74-96], sensing viscosity conductivity of liquids[97], measuring viscoelastic properties of thin films [98], and for biomedical applications, such as glucose sensing[99.100].

Resonant structures have been applied to sensing force, acceleration and pressure[101], and somewhat to chemical vapors[102].  Application to vapor sensing is difficult because the drive side of these structures is in direct contact with the environment.  Most commercially successful resonant sensors have been made of quartz, which has the advantages of piezoelectric coupling and temperature-stable configurations, but the disadvantage of fairly high manufacturing cost. Acoustic wave devices are classified as, the thickness-shear mode (TSM) resonator, the surface acoustic wave (SAW) device, the acoustic plate mode (APM) device, and the flexural plate wave (FPW) device [2], depending on the mode of acoustic wave propagation.

Stuart W. Wenzel and Richard M. White [3] have fabricated a micro sensor that employs ultrasonic Lamb waves propagating in a thin plate supported by a Silicon die. This device is sensitive to many measurands [4]. Therefore it could operate as a microphone, biosensor, chemical vapor or gas detector, scale, pressure sensor, densitometer, radiometer or thermometer for example.

By applying a force to the membrane directly, the membrane tension will increase causing an increase in oscillator frequency. The response to gas pressure can be realized by applying fixed pressure on the other side of the membrane. The changes of membrane tension will cause an oscillator frequency change. If both sides of the membrane are subjected to the same gas pressure, membrane tension will be constant, but loading of the propagating wave will depend upon the pressure. Loading one or both sides of the membrane with a fluid can cause quite large velocity changes and oscillator frequency shifts[5].

Stwart W. Wenzel and Richard M. White have described the use of flexural plate waves traveling in thin composite plates of silicon nitride, zinc oxide and aluminum for gravimetric chemical vapor sensing. Detection in the parts-per-billion range is achievable when the plate is coated with polymer films of poly(dimethyl siloxine) and ethyl cellulose.

C. E. Bradley et al.[5] formed microscopic channels at the interface between and etched glass cap and a silicon micromachined flexural plate wave (FPW) device. The FPW devices excite an acoustic field on the fluid that occupies the channel, and the acoustic field in turn drives fluid flow.

Amy W. Wang et al. [6] developed a new technique which involves the use of ultrasonic waves in gel-coated flexural plate waves sensor.  The flexural plate wave described has a high sensitivity in comparison to other acoustic sensors.  Their experiments are apparently the first in which an acoustic sensor has been used to measured diffusion of liquid in gels [8].

## CHAPTER II

## Device  Applications

In this section we discuss some applications of the Flexural Plate Wave devices.    We begin with gravimetric sensing, which means we will explore experiments based on measurands whose primary effect is to cause a mass change or perturbation of kinetic energy.  Additionally we will examine sensing based on perturbations of the potential energy.

### 2.1    Sensing based on Perturbation of Kinetic Energy

The phase velocity of the Flexural Plate wave device varies with the mass / area of the membrane[4] .

$$v_{phase} = \sqrt{\frac{T_x + D}{mass/area}} \qquad (1)$$

where $T_x$ is the component of in-plane tension in the propagation direction(taken as the x direction)  per unit width in the y direction, perpendicular to the direction of propagation. Explicitly,

$$T_x = \int_0^h \tau_{yy}(\xi)d\xi \qquad\qquad (2)$$

where $\tau_{yy}$ is the y component of tensile stress in the plate and $\xi$ is a variable of integration normal to the plate whose bounding surfaces are at z = 0 and z = h. $D$ is the bending moment of the plate and it is expressed to be:

$$D = \frac{Eh^3}{12(1-\upsilon^2)} \qquad\qquad (3)$$

$E$ is the Young's modules, $h$ is the membrane thickness and $\upsilon$ is Poisson's ratio.

The change of mass/area of a polymer film on the membrane of a delay line oscillator caused by absorption of chemical vapors produces a velocity shift and a directly measured associated frequency shift. Therefore, the flexural plate wave device can be used for chemical vapor sensing individually or as part of a more sophisticated chemical measurement instrument[10].

The density of a liquid is measured by plugging the Flexural Plate Wave sensor into a printed circuit board amplifier. The amplifier frequency is related to the density by the relationship:

$$\frac{1}{f^2} = A + B\rho \qquad\qquad (4)$$

where A has of sec$^2$ and B is a property of the liquid with the unit of sec$^2$ m$^3$/kg. Thus, the Flexural Plate Wave device is a unique solid-state device sensor for measuring the density, viscosity and speed of sound of liquids. Because it is small and potentially inexpensive, it is ideal for in-line density measurement applications[10].

The viscosity of a liquid is measured by applying a signal to one transducer and measuring the received signal at the other transducer. The ratio of these two signals is a function of the liquid's viscosity. This allows one to use the Flexural Plate Wave device to measure liquid viscosity and the dependance of viscosity on temperature[10].

## 2.2    Sensing based on Perturbation of Potential Energy

Differential pressure applied to the plate can change the tension and produce a frequency change, allowing the Flexural Plate Wave device to be used as pressure sensor. We also can produce a frequency shift by directly applying a force on the silicon chip. Thus, the Flexural Plate Wave device can be used for sensing force.

# CHAPTER III

## Theory of Flexural Plate Wave Devices

### 3.1    Why Flexural Plate Waves ?

Figure 3.1 shows that the thickness of the plate is much smaller than the wavelength of the ultrasonic waves propagating in it.

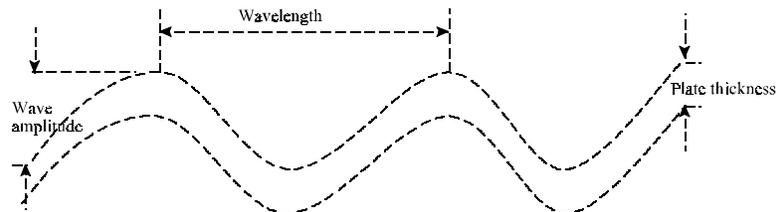

Fig. 3.1 Elastic Plate Wave

In a thin plate wave the phase velocity can be much lower than that of a bulk or surface waves, which leads to a low operating frequency relative to the other ultrasonic sensors, and possibly simpler and less expensive support electronics.

### 3.2    The Study of Flexural Plate Wave

We approach the subject of plate waves with applications to sensing in mind, so it is helpful to reconsider (Figure 1.1)[11] which shows a classification of ultrasonic measurement methods.  Ultrasonic sensors employ the propagation of elastic waves along a path whose characteristics are altered by a measurand (the quantity or physical effect we wish to sense, for example pressure, strain, acceleration or chemical species).  As the characteristics of the path are altered, so are waves properties, such as amplitude, resonant frequency, phase or group velocity, phase, etc.  We measure changes of these properties to get information on the measurands.

Different transduction schemes allow us to excite particular modes and to measure their properties directly or indirectly.  Transduction can be divided into one-port and two-port schemes[2] ( figure 1.1 ).  For two-port schemes, phase velocity and phase velocity shift can be inferred from the phase shift between input and output transducers.  Phase can be measured passively at a constant operating frequency or actively using a feedback or phase shift oscillator.  The dissipation or attenuation can be found from changes of insertion loss (ratio of the output to input voltage or current).  In a one-port scheme, the impedance of the transducer can be measured as a function of frequency to find resonant frequencies of the device and information about energy dissipation wave attenuation.  It is also common to track the natural frequencies actively using an oscillator.

### 3.3     Differential Equation of Motion for Flexural Plate Waves

When a plate is bent, it is stretched at some points and compressed at others.  On the convex side there is extension, which decreases into the plate, until towards the concave side a gradually increasing compression is found.

In the thin plate differential equation of motion, all of the normal components of forces acting on the plate are balanced in order to derive the dispersion relation .  A metal meander-line transducer (MLT) is patterned on the plate (membrane) surface (Figure 3.3.1)[1].  Current lines running across the membrane are positioned at regular intervals on the insulating membrane.

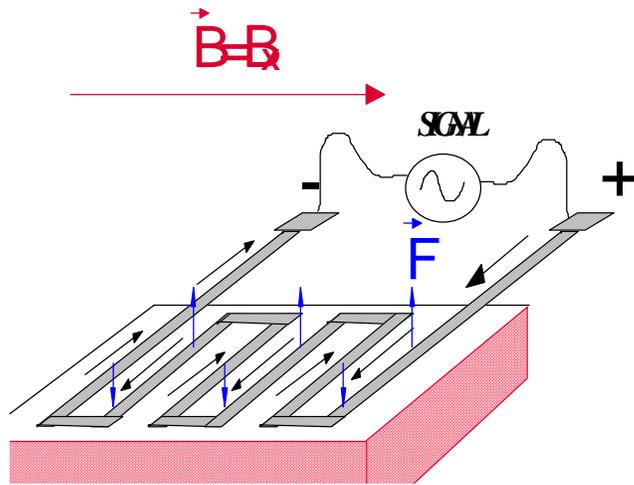

Fig. 3.3.1  One-port magnetically excited Flexural Plate  Wave Device

If we let

$$\vec{f} = \vec{J} \times \vec{B} \qquad\qquad (4)$$

be an external surface-normal Lorentz stress (force per unit area) , which is generated by an

applied static in-plane magnetic field perpendicular to the current direction.   The direction of the

force is determined by the Right Hand Rule (figure3.3.2)[1]

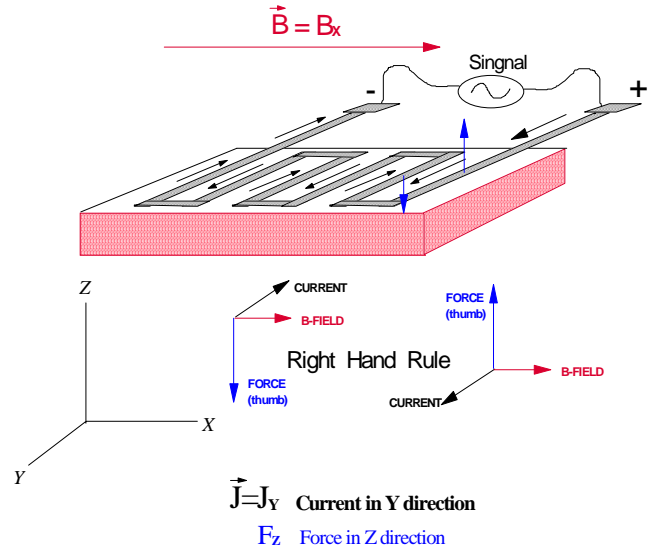

Fig. 3.3.2 Direction of the Lorentz Force
determined by Right Hand Rule

Referring to the geometry of Fig.3.3.2, the total free energy $F_{pl} = \int F\, dV$ of a plate can

be written by[47]

$$F = \frac{Eh^3}{24(1-\nu^2)} \iint \left[ \left( \frac{\partial^2 u}{\partial x^2} + \frac{\partial^2 u}{\partial y^2} \right)^2 + 2(1-\nu)\left\{ \left( \frac{\partial^2 u}{\partial x \partial y} \right)^2 - \frac{\partial^2 u}{\partial x^2} \frac{\partial^2 u}{\partial y^2} \right\} \right] \qquad (5)$$

where $E$, $\nu$, $h$ and u are the membrane Young's modulus, the membrane Poisson ratio, the membrane thickness and the displacement of points on the plate respectively. This form is quadratic in second derivatives because the bending restoring force depends upon curvature differences in the medium. This integral can be divided into two parts and varied separately. The first integral can be written in the form $\int \left( \nabla^2 u \right)^2 df$ where $df = dx\,dy$ is a surface element and

$\nabla^2 = \dfrac{\partial^2}{\partial x^2} + \dfrac{\partial^2}{\partial y^2}$ is the two dimensional Laplacian.  Varying the equation we have

$$\delta \tfrac{1}{2} \int \left( \nabla^2 u \right)^2 df = \int \nabla^2 u \nabla^2 \delta u\, df$$
$$= \int div \left( \nabla^2 u\, grad\, \delta u \right) df - \int grad\, \delta u . grad\, \nabla^2 u\, df \tag{6}$$

The first integral on the right can be transformed into an integral along a closed contour enclosing the plate[47]:

$$\int div \left( \nabla^2 u\, grad\, \delta u \right) df = \oint \nabla^2 u \frac{\partial \delta u}{\partial n} dl \tag{7}$$

In the second integral we use the same transformation to obtain

$$\int grad\, \delta u . grad\, \nabla^2 u\, df = \oint \delta u \frac{\partial \nabla^2 u}{\partial n} dl - \int \delta u \nabla^4 u\, df \tag{8}$$

The second integral may be integrated by parts.  After substituting equations (7) and (8) and after doing the integration by parts for the second integral we get

$$\int \left\{ D \nabla^4 u - P \right\} \delta u\, df = 0 \tag{9}$$

This integral can therefore vanish only if the coefficient of $\delta u$ is zero.

$$D \nabla^4 u - P = 0 \tag{10}$$

where D is the plate bending moment and P is other forces acting on the plate.

Now let us consider that the membrane is perfectly flexible.  The tension $T$ and density $\rho_s$ are uniform throughout the membrane.  In the equilibrium position the membrane is planar and undergoes only very small deflections from its equilibrium position, the only displacements taking place being those perpendicular to the membrane. Let $T$ and $\rho_s$ be perpendicular force per unit length acting on an imaginary line segment embedded in the membrane.

If an elementary section is displaced in direction perpendicular to xy-plane by an amount $\delta u$, then the force acting in this direction is $(T \delta u)$ at length $\delta x$. So the force acting at unit length is

$$\frac{T \delta u}{\delta x} = T \frac{\partial u}{\partial x} \tag{11}$$

and hence the force in perpendicular direction at the edge $x + \delta x$

$$T \frac{\partial u}{\partial x} \Big|_{x+\delta x} \delta y \tag{12}$$

and the force acting at the edge x is

$$-T \frac{\partial u}{\partial x} \Big|_{x} \delta y \qquad \qquad (\text{ in the opposite direction}) \tag{13}$$

So the total force along the edges $x$ and $x + \delta x$ is

$$T \frac{\partial^2 u}{\partial x^2} dy\, dx \tag{14}$$

Similarly, restoring force along the edges at $y$ and $y + \delta y$ is

$$T \frac{\partial^2 u}{\partial y^2} dy\, dx \tag{15}$$

So the total restoring force is given by

$$T\left[\frac{\partial^2 u}{\partial x^2} + \frac{\partial^2 u}{\partial y^2}\right] dy\, dx \tag{16}$$

The deflecting force in this direction accelerates the membrane

$$\rho_s \frac{d^2 u}{dt^2} dx\, dy \tag{17}$$

Newton's second law of motion gives

$$T\left[\frac{\partial^2 u}{\partial x^2} + \frac{\partial^2 u}{\partial y^2}\right] = \rho_s \frac{\partial^2 u}{\partial t^2} \tag{18}$$

Combining equations (10) and (18) and including a dissipative term ($Z_a$ which is the mechanical impedance of air), we will get the equation of motion for a plate flexed by external forces acting on it [47]. We can express the force balance with one differential equation[1]

$$D\nabla^4 u - T\nabla^2 u + Z_a \dot{u} + \rho_s \ddot{u} = \vec{F} \tag{19}$$

Each term in equation (19) has the unit of stress (force per unit area). The first term is due to the flexural rigidity or membrane bending moment $D$ of the plate. The second term comes from the restoring force due to tension $T$ in the plane of the plate. The third term is due to the effect of fluid loading. The mass layer contributes to the fourth term, where $\rho_s$ is the areal mass density. Equation (19) can be solved in two regimes: the linear solutions and the nonlinear solutions.

In this project only linear solutions were examined. In the linear regime, the membrane can be represented by an equivalent circuit model [1] (Figure 3.3.3). That model consists of a parallel LRC resonator that is transformer coupled to the electrical ports. The equivalent circuit elements are given by[1]

$$L_1 = \frac{1}{D k_{mn}^4}$$

$$L_2 = \frac{1}{T k_{mn}^2}$$

(20)

$$C_1 = \rho_s$$

$$R_1 = \frac{\omega}{D \delta k_{mn}^4}$$

where $k_{mn} = \left( k_m^2 + k_n^2 \right)^{1/2}$. $k_m$ and $k_n$ are the transverse and longitudinal wave numbers defined by the membrane boundaries (see section 3.4).

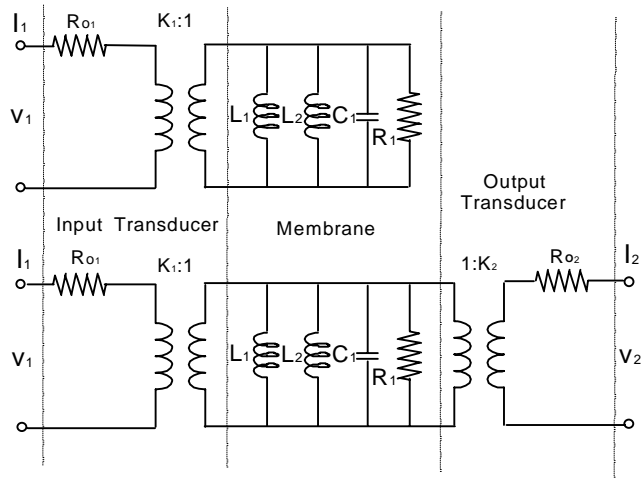

Fig. 3.3.3    Equivalent Circuit Models to depict the resonance of magnetically-excited FPW devices in the vicinity of a membrane resonance when operating in vacuum: (a) one-port device, (b) two-port device[1].

## 3.4    Linear response

The membrane can be excited into a number of eigenmodes, with displacement given as a superposition of normal modes[1]

$$u(x, y, t) = \sum_{m=1}^{\infty} \sum_{n=1}^{\infty} A_{mn} \sin(k_m x) \sin(k_n y) e^{i\omega} \qquad (21)$$

where $A_{mn}$ is the complex displacement amplitude of the $(m,n)$ eigenmode, $k_m = \dfrac{m\,\pi}{l}$ and

$k_n = \dfrac{n\,\pi}{w}$ are obtained by the boundary conditions, where $l$ is the membrane length, $w$ is the

membrane width (figure 7.1), $m$ and $n$ are integers and $\omega$ is the angular frequency.

The eigenmode amplitude $A_{mn}$ can be written[1]

$$A_{mn} = -\frac{2}{j\omega(wl)^{\frac{1}{2}}}\left[\frac{K_{mn}^{(1)}\hat{I}_1 + K_{mn}^{(2)}\hat{I}_2}{Z_{mn}}\right] \tag{22}$$

The mode coupling factor $K_{mn}^{(i)}$ is proportional to the spatial correlation between the

current density and the $(m,n)$ eigenmode displacement [1]

$$K_{mn}^{(i)} = \frac{2B_x}{(wl)^{\frac{1}{2}}\hat{I}_i}\int_0^w\int_0^l \hat{J}_y^{(i)}(x,y)\sin(k_m x)\sin(k_n y)\,dx\,dy \tag{23}$$

where $J_y^{(i)}$ is the amplitude of the current density applied by the $i^{th}$ meander-line transducer.

Integration of $J_y$ over a cross section gives:

$$J_y^{)}(x,z) = I_0\sum_{p=1}^{2N}\delta(x-x_p)(-1)^p\delta(z-z_p) \tag{24}$$

where $N$ is the number of MLT periods.

Using equation (24), equation (23) can be written as:

$$K_{mn} = \frac{2 B_x}{I_0 \sqrt{w\,l}} I_0 \left[ \sum_{p=1}^{2N} (-1)^{p+1} \sin(\frac{m\pi}{l} x_p) \right] \left[ \frac{1 - \cos(\frac{n\pi}{w} y_p)}{\frac{n\pi}{w}} \right] \tag{25}$$

$$K_{mn} = \frac{2 B_x}{\pi} \sqrt{\frac{w}{l}} \left( \frac{1 - (-1)^n}{n} \right) \sum_{p=1}^{2N} (-1)^{p+1} \sin(\frac{m\pi}{l} x_p) \tag{26}$$

The voltage across an meander-line transducer has contribution from ohmic resistance in the line and a back emf due to motion of the conductor in the magnetic field. Therefore it can be written as :

$$\hat{V}_j e^{-i\omega} - \hat{I}_j R_j e^{-i\omega} = -\frac{d}{dt} \iint \vec{B} \cdot d\vec{A} \tag{27}$$

$$\hat{V}_j e^{-i\omega} - \hat{I}_j R_j e^{-i\omega} = -\frac{d}{dt} \iint e^{-i\omega} B_x \hat{u}_p (x, y) dy \tag{28}$$

$$\hat{V}_j = \hat{I}_j R_j + i\omega B_x \sum_{p=1}^{2N} (-1)^{p+1} \sum_{m,n} A_{m,n} \sin(k_m x_p) \times \left[ \frac{1 - (-1)^n}{n\pi} \right] w \tag{29}$$

$$\hat{V}_j = \hat{I}_j R_j + i\omega B_x \sum_{lines} \int_0^w \hat{u}_p (x, y) dy \tag{30}$$

The alternate lines have currents in the same direction, and therefore we have the same sign for the induced voltage. Adjacent lines have currents in the opposite direction, and therefore induce an opposite sign for the voltage. Equation (30) can be written as

$$\hat{V}_j = \hat{I}_j R_j + i\omega B_x \left[\frac{w}{2B_x}\right]\left[\frac{l}{w}\right]^{\frac{1}{2}} \sum_{m,n} A_{mn} k_{mn} \tag{31}$$

By simplifying, equation (31) becomes

$$\hat{V}_j = \hat{I}_j R_j + \frac{i\omega}{2}(lw)^{\frac{1}{2}} \sum_{m,n} A_{mn} k_{mn} \tag{32}$$

## CHAPTER IV

## Silicon Carbide and Silicon Nitride Films

### 4. 1    Silicon Carbide Films

Silicon Carbide was discovered by Berzelius in 1824 while attempting to produce synthetic diamond. Silicon Carbide is an attractive candidate as a semiconductor material for high-temperature and high-power applications because of its large bandgap (2.3-3.3 eV), good carrier mobility, and excellent thermal and chemical stability. The fabrication technology of Silicon Carbide electronic devices moreover is based on the Silicon microelectronics industry. A change of technology from Silicon to Silicon Carbide will revolutionize the power industry (Figure 4.1) because it will then be economically feasible to use power electronics to a much larger extent than today (for example to control engines). Tong et al[11] evaluated both epitaxial

Silicon Carbide and sputtered amorphous films as a new micro mechanical material. One

property, which makes the Silicon Carbide films particularly attractive for micro machining is the

fact that these films can easily be patterned by dry etching using an Al mask. Patterned Silicon

Carbide films can further be used as passivation layers in the micro machining of the underlying

Silicon substrate (Silicon Carbide can withstand KOH, HF and TMAH etching. Marc Madou[14] et al used single crystal Silicon Carbide to fabricate high temperature pH sensors.

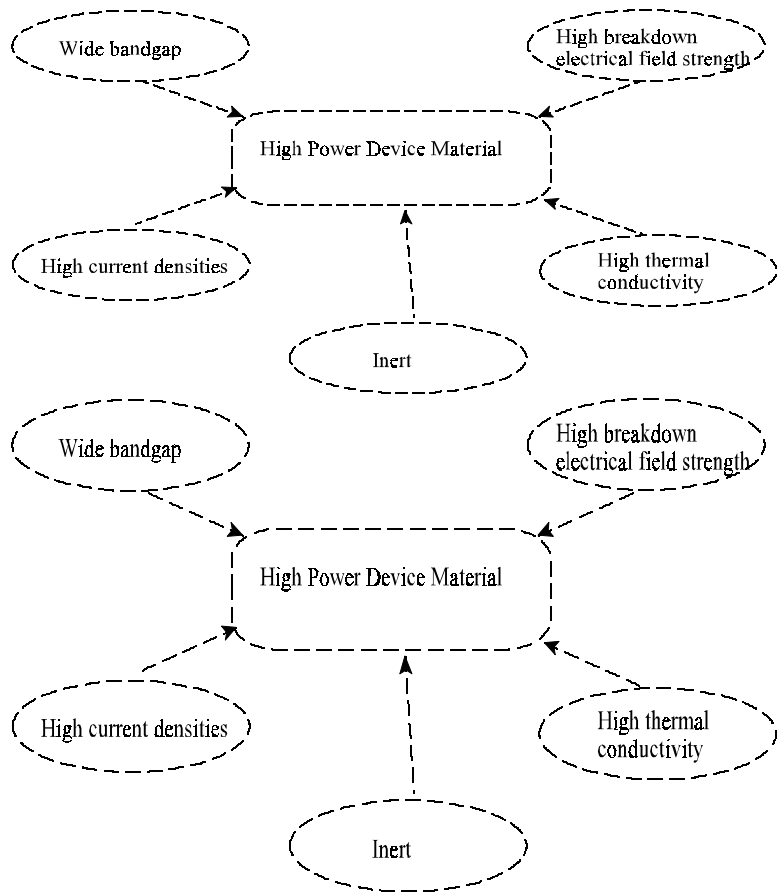

Fig. 4.1 Application of Silicon Carbide

## 4.2    Silicon Nitride Films

Silicon Nitride is a commonly used material in microcircuit and micro sensor fabrication due to its many superior chemical, electrical, optical, and mechanical properties. Silicon Nitride is also a hard material, but it can be etched by HF. Some applications of Silicon Nitride are optical waveguides, insulators ( high dielectric strength), mechanical protection layers, and etch masks. The published characteristics of the Silicon Nitride films will be used in this study.

After fabrication of the Silicon Carbide device and characterization of the device, the results will be compared with the published results of typical Silicon Nitride devices used as transducers.  Table 4.2 shows the properties of Silicon Carbide (SiC) and Silicon Nitride (SiN).

**Table 4.2  Properties of Silicon Carbide and Silicon Nitride**

|  | Silicon Carbide | Silicon Nitride |
|---|---|---|
| Density (g/cm$^3$) | 3.21-3.27 | 2.9-3.2 |
| Dielectric constant | 9.7 | 6-7 |
| Resistivity ($\Omega$cm) | doping dependance | $10^{16}$ |
| Refractive index | 2.69 | 2.01 |
| Energy gap eV | 2.3-3.3 | 5 |
| Dielectric strength ($10^6$ V/cm) | $\geq 1.5$ | 10 |
| Etch rate in con.HF | N/A | 200 Å/min |
| Etch rate in BHF | N/A | 5-10 Å/min |
| Residual stress ($10^9$ dyn/cm$^2$ ) |  | 1 T |
| Poisson ratio | 0.14 | 0.27 |
| Young's modulus | 392 GPa | 270 GPa |

# CHAPTER V

## Device Fabrication

### 5.1    Flexural Plate Wave Device Mask Layout

Once a suitable mask (Figure 5.1a and 5.1b) set has been designed and obtained, the next step towards obtaining the Flexural Plate Wave device is the fabrication process.  The first phase has to do with the preparation of the sample for the fabrication process.  The Chemical Vapor Deposition, the Oxidation Process, the Metallization processes, the Photolithographic and the Etch back  follow the first phase.

Upon the completion of fabrication the device must be mounted in a suitable fixture to evaluate its performance.  Each of these steps must be optimized in order to maximize the device performance.

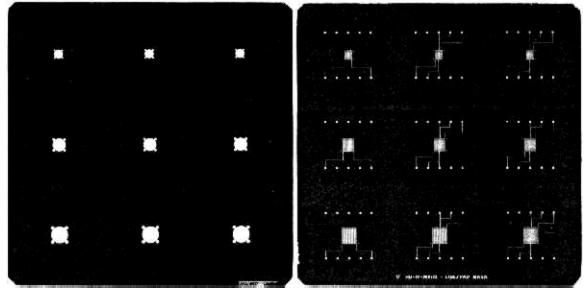

| Fig. 5.1a Backside mask for membrane release | Fig.5.1b Top mask for metal meander line transducer |

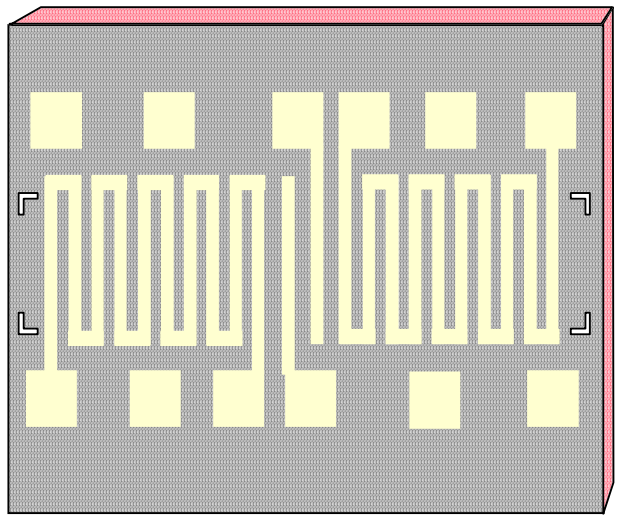

Fig. 5.2 Meander line metal and bond pad mask for
2-Port Device

This chapter discusses each process in detail, and Figure.5.3 shows the process flow diagram used for the fabrication of the Flexural Plate Wave device.

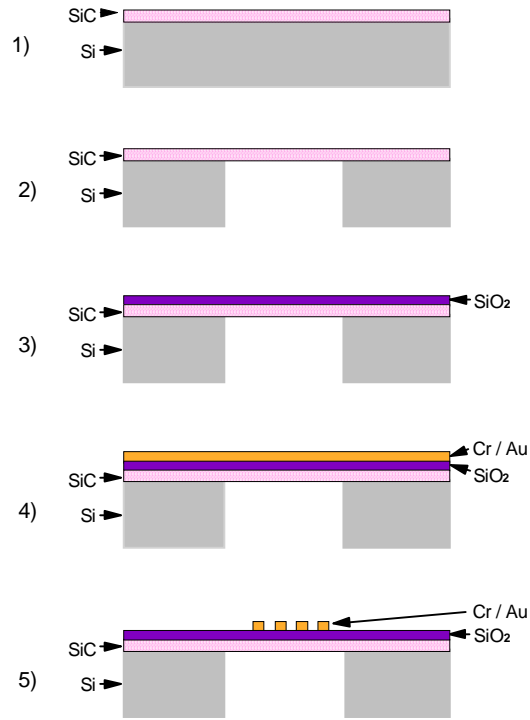

Fig. 5.3    (1) Si wafer with CVD SiC layer, (2) free-standing SiC membrane, (3) thermally grown $SiO_2$ layer on top of the SiC layer for insolation, (4) Cr/Au metallization layer, (5) MLT has been patterned on the front side and aligned with the backside opening.

## 5.1    Sample Preparation

We start with a single polished, dopant boron, p-type, diameter 50.67 mm, thickness 254 μm, resistivity 52.7 ohm-cm wafer. Upon exposure to oxygen, the surface of a silicon wafer oxidizes to form silicon dioxide. This native silicon dioxide is a high quality electrical insulator.

Therefore an oxide etch is performed using Hydrofluoric acid (HF) solution. Then the wafer is rinsed in De-Ionized water (DI) and then blown dry with Nitrogen gas ($N_2$). The wafer is then ready for the growth of Silicon Carbide using Chemical Vapor Deposition (CVD).

**5.2     Chemical Vapor Deposition of Silicon Carbide (SiC)**

Chemical vapor deposition forms thin films on the surface of a substrate by thermal deposition and/or reaction of gaseous compounds. The desired material is deposited directly from the gas phase onto the surface of the substrate[15]. The CVD reactor used for this project is a vertical CVD reactor. The growth condition are the following: Temperature: 1100°C, Pressure: 200 Torr, Gas/flow rate: Silane /25sccm (5%); Propane/15sccm(20%); Shroud/28sl per min, Time: 10 minutes. With these conditions, a thickness between 1-1.5 $\mu m$ can be obtained[14]. Once a CVD silicon carbide coated wafer is obtained (with a thin deposition of SiC on the wafer back side), the next step is the growth of silicon dioxide.

## 5.3　Oxidation of Silicon Carbide

Silicon Dioxide $SiO_2$ was grown thermally on the Silicon Carbide Membrane.  For example the growth were done at 1200 degrees Celsius for 4 hours.   The thickness of the $SiO_2$ layer was about 0.5 microns.  The $SiO_2$ thickness is limited to a few microns due to the fact that SiC does not oxidize easily.   The $SiO_2$ layer was used for electronic isolation.

## 5.4　Metal deposition

Physical evaporation represents one of the oldest metal films deposition techniques.  Evaporation is based on the boiling off of a heated material onto a substrate in a vacuum.  From thermodynamic considerations, the number of molecules leaving a unit area of evaporant per second is given by:

$$N = N_0 \exp - \left( \frac{\Phi_e}{kT} \right)$$

where $N_0$ may be a slowly varying function of temperature, $T$, and $\Phi_e$ is the activation energy required to evaporate one molecule of the material[14].  For this project 150 Å of Chrome (Cr) and 5000Å of Gold (Au) are heated to the point of vaporization, and then evaporate to form a thin film covering the surface of the silicon dioxide ($SiO_2$).  The base pressure before evaporation was 2- 4 x $10^{-7}$ Torr. The Cr layer is needed to have a good adhesion.

**5.5    Photolithography**

The photolithographic process is one of the most critical steps in the device fabrication process. It encompasses all the steps involved in transferring a pattern from a mask to the surface of a wafer.   As the first steps in the lithography process itself, a thin layer of an organic polymer, a photoresist, sensitive to ultraviolet radiation, is deposited on the surface of the Cr-Au layer. The photoresist is dispensed from a viscous solution of the polymer onto the wafer laying on a wafer plate in a resist spinner.  A vacuum chuck holds the wafer in place.  The wafer is then spun at a speed of 4000 rpm to make a uniform film.  The speed depends on the viscosity and the required film thickness.   At that speed, centrifugal forces cause the solution to flow to the edges where it builds up.  The resulting polymer thickness, *h*, is a function of spin speed, solution concentration, and molecular weight.  The expression for *h* is given by[14]

$$h = \frac{K C^{\beta} \eta^{\gamma}}{\omega^{\alpha}}$$

with $K$  a calibration constant, $C$ the polymer concentration in grams per 100 ml solution, $\eta$  the intrinsic viscosity, and $\omega$ the number of rotation per minute (rpm). Once the exponential factors ( $\alpha, \beta, \gamma$ ) have been determined, the above equation can be used to predict the thickness of the film to be spun for various molecular weights and solution concentrations of a given polymer.  For this project the photoresist used was the Shipley Inc. S1808[TM] photoresist, which is 0.8 μm thick when spun at 4000 rpm for 30 seconds.  The quality of the resist coating determines the density of defects transferred to the device under construction.  Therefore the spinning process is of primary importance to the effectiveness of pattern transfer.  The application of too much resist results in edge covering and  hillocks, reducing manufacturing yield.  After spin coating, the resist still contains up to 15% solvent and may contain built in stresses. The wafer is therefore prebaked (soft baked) at 95$^\circ$C for 20 minutes to remove solvents and stress to promote adhesion of the resist layer to the wafer. After soft baking, the photoresist is ready for mask alignment and exposure.

A photomask, a glass plate with a patterned emulsion or metal film on one side, [figure.4.1] is placed over the wafer. With manual alignment equipment, the wafer is held on a vacuum chunk and carefully moved into position below the mask. Following alignment, the photoresist is exposed through the mask with high intensity ultraviolet light. The resist is exposed wherever the Cr-Au layer is to be removed. The exposure time was 12 seconds.

Next, the photoresist is developed using a Shipley Microposit 351 $^{TM}$ developer which was mixed 4:1 DI water to 351 concentrate in order to obtain a high resolution. The development time was 1 minute. Then the wafer is rinsed immediately in DI water and then blown dry. Any resist which has been exposed to the UV light is washed away, leaving the Cr-Au layer in the exposed areas. Following exposure and development, a postbaking step is needed at 120 $^\circ$C for 20 minutes to harden the resist and improve adhesion to the substrate.

## 5.6     Metal Etch process

Following the photolithography process, the wafers were taken through a metal etch process which involved soaking the wafer in acetone to remove the photoresist, then placed in an Au etch solution to remove the unwanted Au metal, and finally in a chrome etch solution to remove the unwanted chrome metal. After the metal etch process the devices were inspected for broken meander lines and short-circuits. Then the devices were ready for testing.

# CHAPTER VI

## Device Characterization

### 6.1 Device Characterization

The electrical network can be described by a number of parameter sets. These parameters relate total voltages and total currents at each of the network ports. The scattering parameters are network variables. They include the Z-parameters, H-parameters and Y-parameters. The only difference between the parameter sets is the choice of independent and dependent variables. The parameters are the constants used to relate these variables. The determination of the parameters requires the use of open- and short-circuits in some ports. At low frequencies open- and short-circuit conditions are easily attained. At higher frequencies, these parameters are difficult to measure accurately because the required open- and short-circuit tests are difficult to achieve over a broadband range of microwave frequencies.

Scattering parameters (S-parameters) are used in the characterization of acoustic wave devices. For this experiment, a Hewlett-Packard HP 3577A-network analyzer with the HP 35677A S-parameters test set (100KHz to 200MHz) were used. [figures 6.1 and 6.2]. The scattering parameters (S-parameters) are very important at high frequencies. The S-parameters are tensor quantities that give phase related information about the device input / output characteristics. These are frequency dependent quantities that yield information about the reflection / transmission coefficients, the voltage standing wave ratio (VSWR) and the impedance.

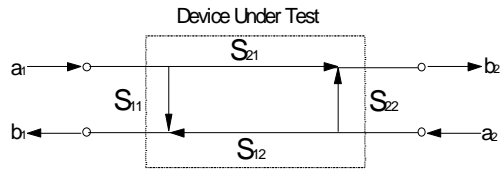

Fig. 6.1   Scattering Parameters

The scattering matrix [S],  for a two-port network can be written as

$$\begin{bmatrix} b_1 \\ b_2 \end{bmatrix} = \begin{bmatrix} S_{11} & S_{12} \\ S_{21} & S_{22} \end{bmatrix} \begin{bmatrix} a_1 \\ a_2 \end{bmatrix}$$ (33 )

For $S_{11}$ , we terminate the output port of the network analyzer and measure the ratio  $b_1$  to  $a_1$

(eq. 34). Terminating the output port in an impedance equal to the characteristic impedance of the

device is equivalent to setting $a_2 = 0$ , since a travelling wave incident on this load will be totally

absorbed.   $S_{11}$ is the input reflection coefficient of the network.  Under the same condition, we

can measure $S_{21}$ , the foward transmission through the network.  This is the ratio of $b_2$  to  $a_1$

(eq.35), which could be either the gain of an amplifier or the attenuation of a passive network.

$$S_{11} = \frac{b_1}{a_1} \Big|_{a_2 = 0}$$ (34)

$$S_{21} = \frac{b_2}{a_1} \Big|_{a_2 = 0}$$ (35)

By terminating the input side of the network, we set $a_1 = 0$. $S_{22}$, the output reflection

coefficient, and $S_{12}$, the reverse transmission coefficient, can be measured (eqs. 36 and 37)

$$S_{22} = \frac{b_2}{a_2} \Big|_{a_1 = 0} \tag{36}$$

$$S_{12} = \frac{b_1}{a_2} \Big|_{a_1 = 0} \tag{37}$$

The relationship between the traveling waves is easily seen. We have $a_1$ incident on the network.

Part of it transmits through the network to become part of $b_2$. Part of it is reflected to become part of $b_1$. Meanwhile, the $a_2$ wave entering port two is transmitted through the network to become part of $b_1$ as well as being reflected from port two as as part of $b_2$.

## 6.2    Experimental Set up

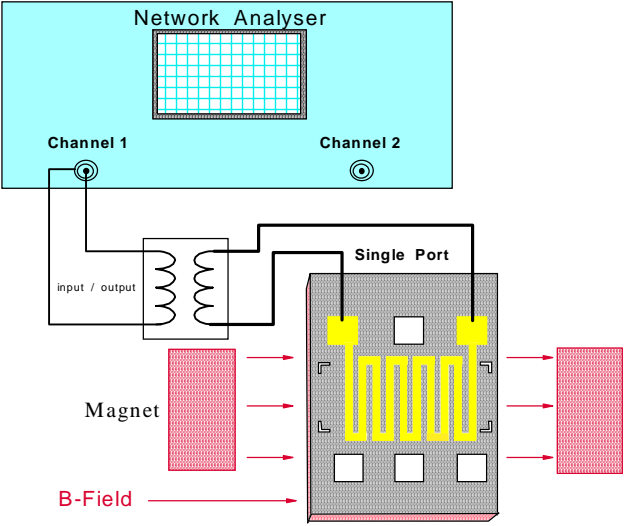

Fig.  6.2.1    Experimental Setup for One-port
FPW Device

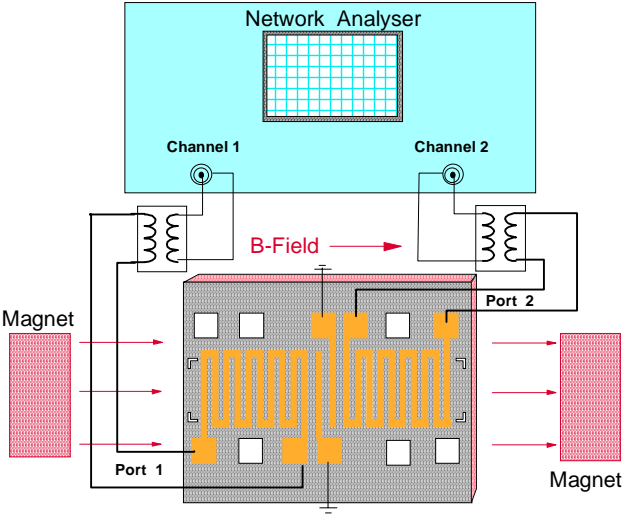

Fig.  6.2.2    Experimental Setup for Two-port
FPW Device

# Chapter VII

## Results and discussion

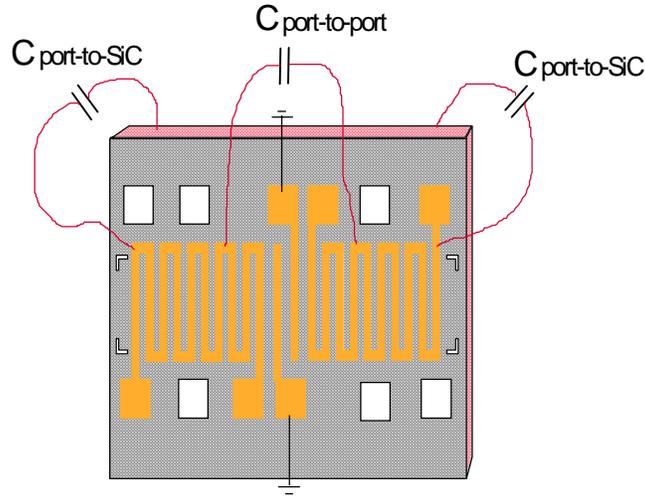

Fig 7.1   Layout of the two-port FPW device

From the geometrical dimension of the FWP device, the length, width and spacing between meander lines, bounds for the capacitance and the inductance can be calculated. Using an expression from the electromagnetism for the capacitance between two parallel conducting plates, sandwiched by a dielectric, a value for the capacitance for 10 meander lines for each port can be obtained. The capacitance of a single meander line of area A and thickness d of the SiO2 dielectric with dielectric constant $\varepsilon$ can be estimated as $C = \varepsilon \frac{A}{d}$. The width and spacing between the meander lines are 74.74e$^{-6}$ m. The 10 meander widths and 9 spaces are together 0.00142 m for the total width. From the physical dimensions of the device, a direct (port-to-SiC) capacitance and a transfer (port-to-port) capacitance is determined as $1.11e^{-8}$ Farads and $4.5e^{-11}$ Farads, respectively.

For the internal inductance, the contribution that two adjacent meander lines contribute is obtained from an expression the gives the inductance for parallel conducting lines connected on one end [1]. Since there are 10 meander lines at one port, there are 5 pairs of conducting lines. An estimate for the total inductance contributed by the 10 meander lines would be a factor of 5 times the inductance contributed by a single pair. The inductance for a conducting pair is

$$L = \frac{l\mu_0}{\pi} \ln\left[\frac{d}{R} + \frac{1}{4} - \frac{d}{l}\right]$$ , where the dimensions are; d = 74.74e-6 m, R = d/4, spacing = d,

and l = 0.00142 m. Therefore, $L = 4.07e^{-9}$ Henries, and an estimate for the total inductance is 5 x L = $2.0e^{-8}$ Henries.

From the transfer capacitance of $4.5e^{-7}$ Farad and total inductance, an electrical resonance of about 65 MHz is obtained, while the mechanical resonance frequencies are in the range of 980 KHz. Thus, the internal capacitance and self inductance of the meander lines will not significantly alter the mechanical resonance of the device.

Since the two-port devices have very low impedance 50 Ohms, the port-to-port capacitance contribute an insignificant amount of coupling at the operating frequency of the device. Moreover, the center line grounding offers additional shielding, reducing further the port-to-port capacitance. Similarly, the mutual inductance is also negligible. Therefore, the characteristic waveforms of the device is due primarily to geometrical-mechanical characteristics, not capacitance.

**Table 7.1   Geometrical factors**

| device | MLT | h (m) | w (m) | l (m) | delta (m) |
|--------|-----|-------|-------|-------|-----------|
| one-port | 10 | 1.5 x10$^{-6}$ | 3.71 x 10$^{-3}$ | 3.54 x 10$^{-3}$ | 141 x 10$^{-6}$ |
| two-port #1 | 10 | 1.5 x10$^{-6}$ | 3.11 x 10$^{-3}$ | 3.89 x 10$^{-3}$ | 148 x 10$^{-6}$ |
| two-port #2 | 8 | 1.5 x10$^{-6}$ | 2.42 x 10$^{-3}$ | 3.82 x 10$^{-3}$ | 111.1 x 10$^{-6}$ |

The geometrical factors of each devices are given in Table 7.1, where MLT is the number of meander lines transducer in port 1, h is the membrane thickness in meters w, is the membrane width, delta is the meander spacing and l is the membrane length given by equation (38)

$$l = x_2 (mn) + delf \tag{38}$$

$$x_2 (r_-) = x_1 (mn) + (r + 2) delta \tag{39}$$

$$x_1 (r_-) = delo + (r - 1) delta \tag{40}$$

where delf is the membrane right boundary in m, $r$ which is the meander line position and delo is the membrane left boundary are given by table 7.2

**Table 7.2   Geometrical factors**

| device | one-port | two-port#1 | two-port#2 |
|--------|----------|------------|------------|
| delf (m) | 1085 x 10$^{-6}$ | 565 x 10$^{-6}$ | 1040 x 10$^{-6}$ |
| delo (m) | 1190 x 10$^{-6}$ | 245 x 10$^{-6}$ | 895 x 10$^{-6}$ |

The actual shapes and dimensions of my devices are shown in the following figures

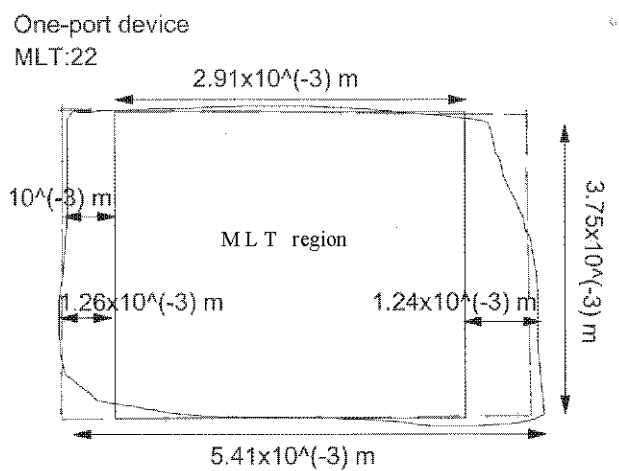

Fig.7.2  Two-port FPW device (device#1)

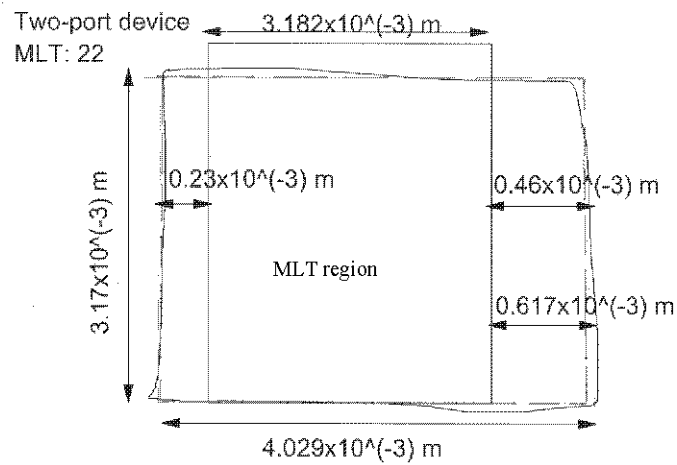

Fig.7.3  Two-port FPW device (device#2)

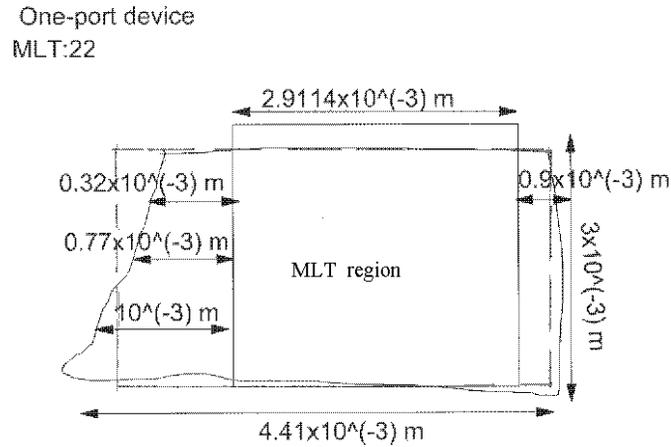

Fig.7.4   One-port FPW device

The mechanical impedance factors of the devices are:  Young's modulus $E$ = 3.92 x $10^{11}$ N/m$^2$, membrane loss tangent $\sigma$ = 0.14, silicon carbide volume density $\rho_1$ = 3210 kg/m$^3$, chrome volume density $\rho_2$ = 7200 kg/m$^3$, chrome thickness $h_2$ = 0.015 x $10^{-6}$m, gold volume density $\rho_3$ = 19300 kg/m$^3$ and gold thickness $h_3$ = 0.5 x $10^{-6}$m.

The fluid properties are:  Air volume density $\rho_f$ = 1.2 kg/m$^3$, the viscosity of air $\eta_f$ = 2x $10^{-5}$ kg/ms and the longitudinal speed of sound $c_l$ = 346 m/s.  These mechanical factors help us to calculate the membrane areal mass density using equation (41)[1].

$$\rho_s = \rho_1 h \ + 0.82\left(\rho_2 h_2 + \rho_3 h_3\right) \tag{41}$$

The Electrical Impedance factors are given by the following equations:

$$Z_{11}(\omega, T, B, C, L) = \frac{1}{Z_0}\left(R_0 + \frac{1}{i\omega C_d} + i\omega L_d + \sum\left[\sum\frac{\left[K_1(B, l, w, m, n)\right]^2}{Z_{mn}(\omega, T, Z_a(\omega, T, m), m, n)}\right]\right) \tag{42}$$

$$Z_{12}(\omega, T, B, C, L) = \frac{1}{Z_0}\left(\frac{1}{i\omega C_t} + \sum\left[\sum\frac{\left[K_1(B, l, w, m, n)K_2(B, l, w, m, n)\right]}{Z_{mn}(\omega, T, Z_a(\omega, T, m), m, n)}\right]\right) \tag{43}$$

$$Z_{21}(\omega, T, B, C, L) = \frac{1}{Z_0}\left(\frac{1}{i\omega C_t} + \sum\left[\sum\frac{\left[K_1(B, l, w, m, n)K_2(B, l, w, m, n)\right]}{Z_{mn}(\omega, T, Z_a(\omega, T, m), m, n)}\right]\right) \tag{44}$$

$$Z_{22}(\omega, T, B, C, L) = \frac{1}{Z_0}\left(R_0 + \frac{1}{i\omega C_d} + i\omega L_d + \sum\left[\sum\frac{\left[K_2(B, l, w, m, n)\right]^2}{Z_{mn}(\omega, T, Z_a(\omega, T, m), m, n)}\right]\right) \tag{45}$$

where $\omega$ is the angular frequency of the device, T is the tension in the Silicon Carbide film, B is the magnetic field strength, $C_d$, $L_d$ and $C_t$ are the direct and transfer capacitance and inductance associated with the given impedance factors. $Z_{mn}$ is the mechanical impedance ( ratio of surface normal stress to particle velocity) associated with excited the mode [1] :

$$Z_{mn} = \frac{Dk_{mn}^4}{i\omega} + \frac{Tk_{mn}^2}{i\omega} + i\omega\rho_s + \frac{D\sigma k_{mn}^4}{\omega} + Z_a \tag{46}$$

$Z_a$ is defined to be the mechanical impedance of the fluid ( air).

$$Z_a(\omega, T, m) = \frac{-2i\omega\rho_f s_m(\omega, T, m)}{k_m^2 - q_m(T, m) s_m(\omega, T, m)} \tag{47}$$

$$\sigma(T) = \sqrt{\frac{T + D}{\rho_s}} \tag{48}$$

$$c_t^2(\omega) = \frac{i\omega\eta_f}{\rho_f} \tag{49}$$

$$q_m(T, m) = k_m \sqrt{1 - \left(\frac{v}{c_l}\right)^2} \tag{50}$$

$$s_m(\omega, T, m) = k_m \sqrt{1 - \frac{v^2}{c_t^2}} \tag{51}$$

The scattering parameters are determined to be:

$$S_{11}(\omega, T, B, C, L) = \frac{\left((Z_{11} - 1)(Z_{22} + 1) - Z_{12}Z_{21}\right)}{\left((Z_{11} + 1)(Z_{22} + 1) - Z_{12}Z_{21}\right)} \tag{52}$$

$$S_{21}(\omega, T, B, C, L) = \frac{2Z_{21}}{\left((Z_{11} + 1)(Z_{22} + 1) - Z_{12}Z_{21}\right)} \tag{53}$$

The best value for the membrane internal stress ( tension ) was determined graphically. The plots given by figures 7.5, 7.6 and 7.7 were used to give an established range of possible tension values for each Flexural Plate Wave devices.

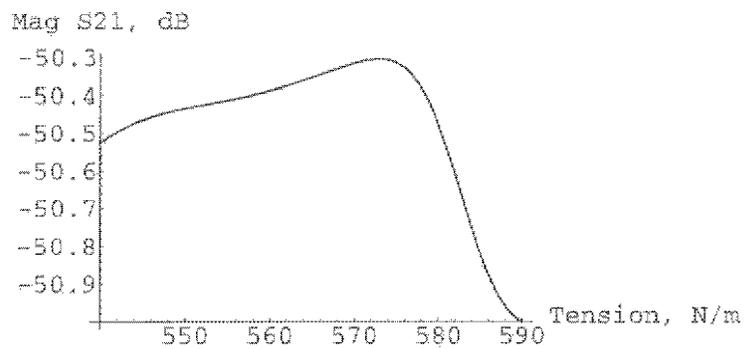

Fig.7.5   Mag S$_{21}$ versus Tension of Two-port FPW device (device #1)

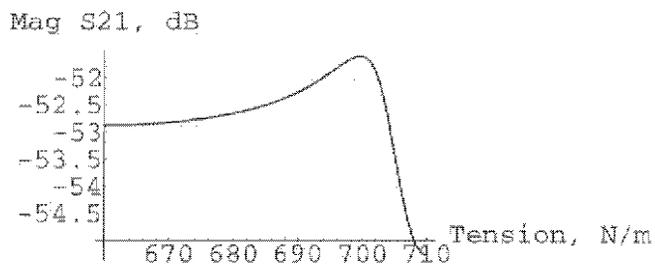

Fig.7.6   Mag S$_{21}$ versus Tension of Two-port FPW device (device #2)

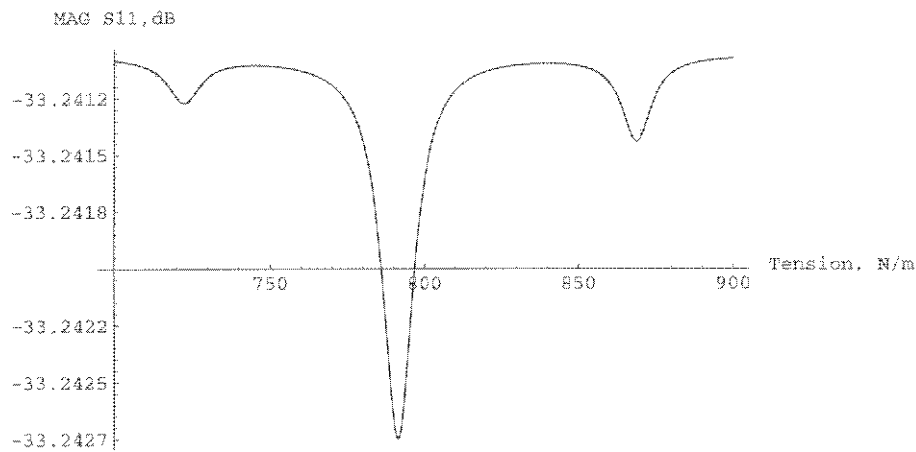

Fig.7.7 Mag S$_{11}$ versus Tension of One-port FPW device

The magnitude of the S$_{21}$ response characteristic for the two port device is compared for the experimental and theoretical cases (Figures 7.8, 7.9, 7.10, and 7.11). The response curves plots for the experimental and theoretical cases are examined for phase and amplitude characteristics, where the magnitude shows minimum and maximum peaks around the characteristic resonance frequency. The phase plots exhibit the locations of the resonance frequencies and the corresponding magnitude values. Tables 7.3 and 7.4 show the comparison between the plotted experimental data and the theory.

**Table 7.3** **Comparison between the plotted experimental data and the theory for two-port FPW device ( device #1)**

| Experiment | | | Theory | | |
|---|---|---|---|---|---|
| Frequency (Hz) | Magnitude (dB) | Phase (deg) | Frequency (Hz) | Magnitude (dB) | Phase (deg) |
| 767000 | -56.5 min | | 786000 | -55.6 min | |
| 780000 | -53.5 mid | 78 | 788000 | -53.3 mid | 72 |
| 790000 | -51.0 max | | 790000 | -51.5 max | |
| | | | | | |
| 796000 | -54.0 min | | 819000 | -52.3 max | |
| 805000 | -53.0 mid | 55 | 820000 | -52.7 mid | 41 |
| 810000 | -52.0 max | | 821000 | -53.0 min | |

**Table 7.4**     **Comparison between the plotted experimental data and the theory for two-port FPW device ( device #2)**

| Experiment | | | Theory | | |
|---|---|---|---|---|---|
| Frequency (Hz) | Magnitude (dB) | Phase (deg) | Frequency (Hz) | Magnitude (dB) | Phase (deg) |
| 991250 | -53.3 min | | 1000000 | -53.0 | |
| 995000 | -52.4 mid | 83 | 1005000 | -52.8 | 75 |
| 999000 | -52.5 max | | 1010000 | -52.6 | |
| | | | | | |
| 1010000 | -52.7 min | | 1037000 | -53.0 | |
| 1025000 | -51.0 mid | 84 | 1041000 | -52.4 | 68 |
| 1047500 | -50.25 max | | 1045000 | -51.9 | |
| 1046000 | -51.5 mid | | | | |

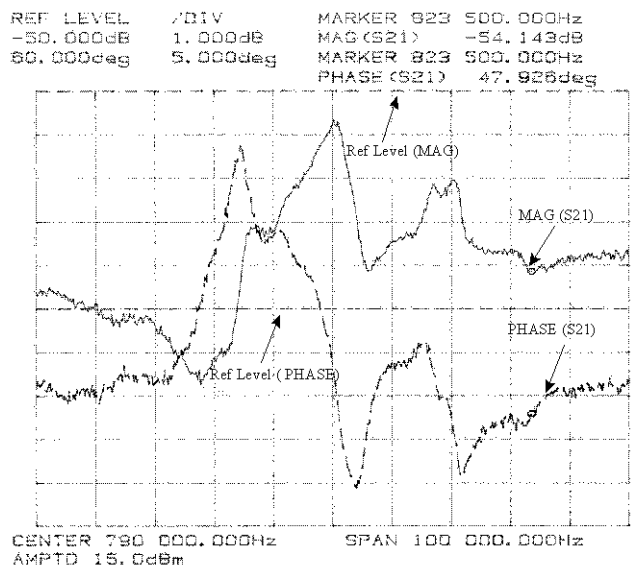

Fig. 7.8     Transmission response S21 (amplitude and phase) of the Two-port FPW Device (Device **#1**, $B_x$ = **9 KG**)

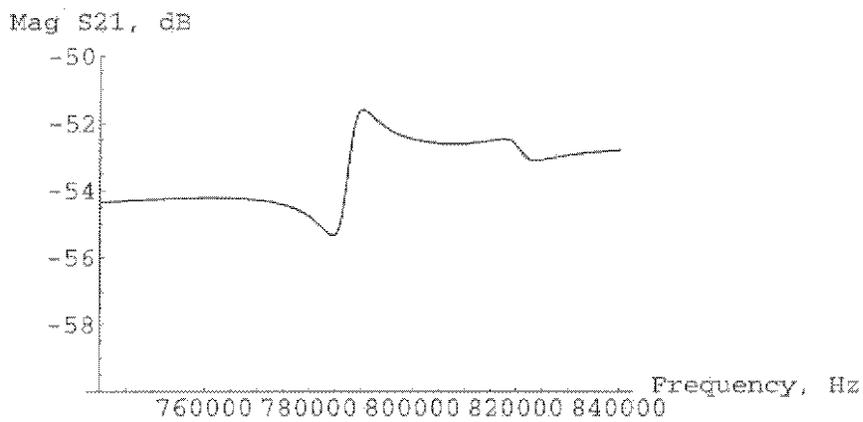

Fig. 7.9.a    $S_{21}$ Magnitude vs Frequency Curve (device #1)

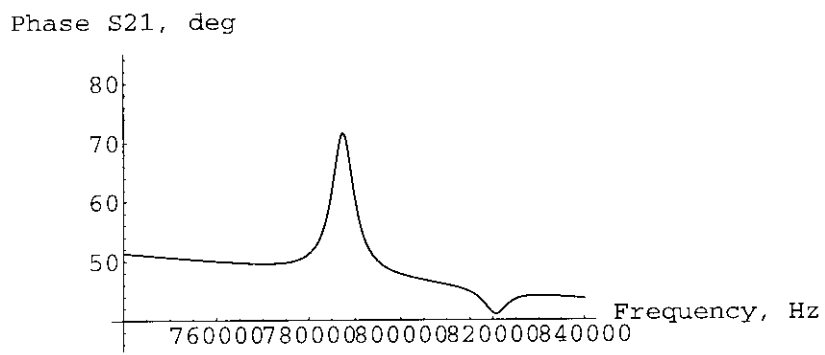

Fig. 7.9.b $S_{21}$ Phase vs Frequency Curve (device #1)

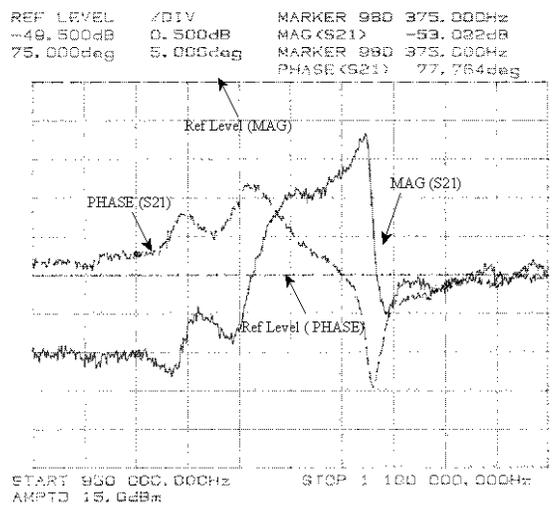

Fig. 7.10    Transmission response (amplitude and
phase) of the Two-port FPW Device

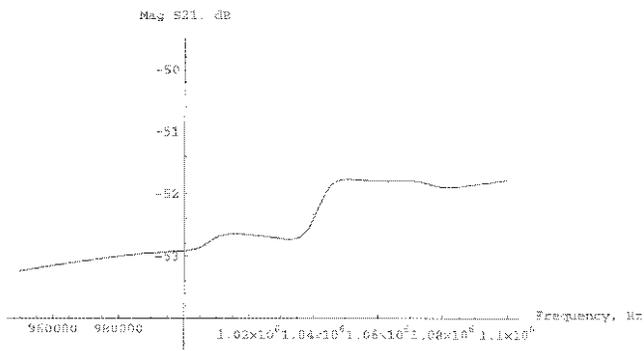

Fig. 7.11.a $S_{21}$ Magnitude vs Frequency Curve (device #2)

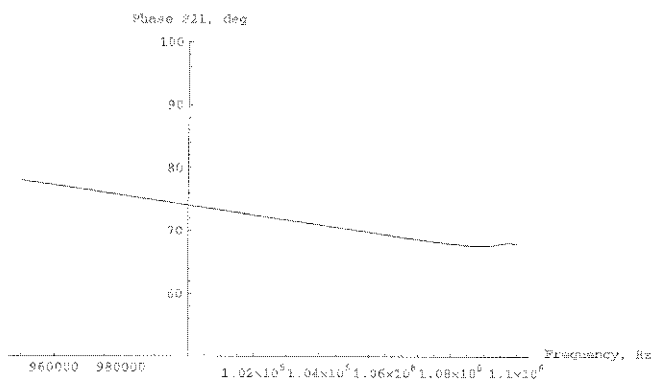

Fig. 7.11.b    $S_{21}$ Phase vs Frequency Curve (device #2)

The response characteristics for the one port device is compared for the experimental and theoretical cases (figures 7.12, 7.13, and 7.14). Here, just as was done with the two port device, the experimental and theoretical response curves for the one port device are examined for the phase and amplitude characteristics. The magnitude exhibits minimum and maximum peaks as a function of frequency. The amplitude as a function of frequency curve exhibits the characteristic resonance frequency for different magnetic field strengths, namely, at the field strengths of 0.3T, 0.46T, 0.6T and 0.7T (figures 7.12 and 7.14a). The location of the resonance frequency is clearly seen at the corresponding magnitude values at the different field. strengths. For the one port device (as well as was found for the two-port devices), the resonance frequency appears to be virtually independent of the field strength.

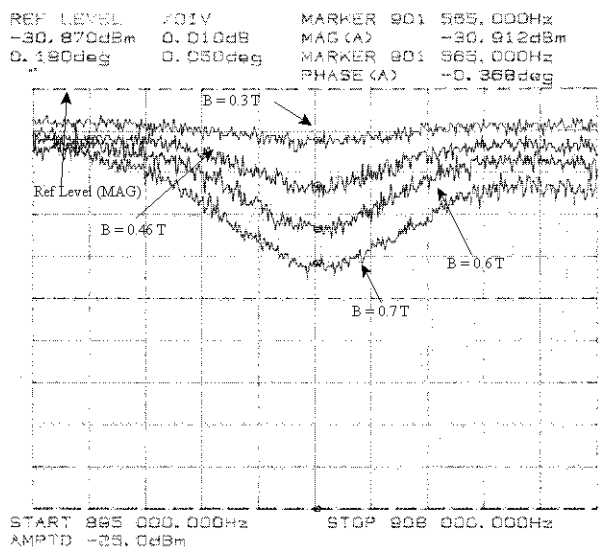

Fig. 7.12    Reflection response $S_{11}$ (magnitude) of a
One-port flexural plate wave device for
different values of the magnetic field

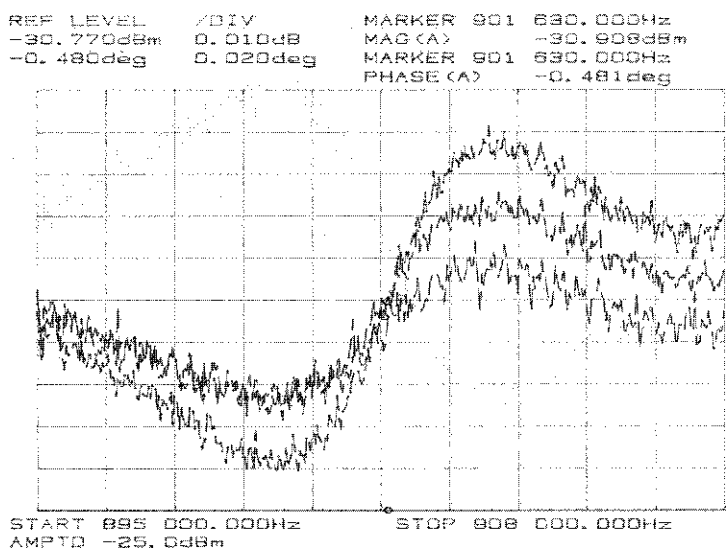

Fig. 7.13    Reflection response S11 (phase) of a One-
port device   for different values of the
magnetic field

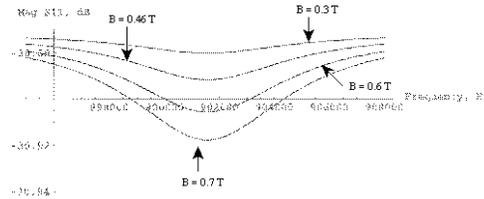

Fig. 7.14.a    $S_{11}$ Magnitude vs Frequency Curve for
different values of the magnetic field (one-
port FPW device)

The magnitude and phase characteristic as function of frequency illustrated in table 7.5

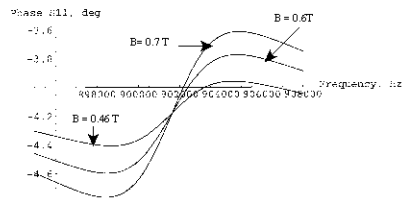

Fig. 7.14.b    $S_{11}$ Phase vs Frequency Curve (one-port
FPW device)

show that the physical device behaves as predicted in the theory. The geometrical configuration and imperfections in the fabrication of the device itself are responsible for any deviations of the physical device response characteristics from the theory. Any imperfections in the fabrication, such as non-rectangular edges, will cause wave reflections from the edges, deviating from the theoretical and ideal device.

**Table 7.5    Comparison between the plotted experimental data and the theory for one-port FPW device (B = 0.7 T )**

| Experiment | | | Theory | | |
|---|---|---|---|---|---|
| Frequency (Hz) | Magnitude (dB) | Phase (deg) | Frequency (Hz) | Magnitude (dB) | Phase (deg) |
| 901565 | -30.89 | -0.481 | 901750 | -30.91 | -0.425 |

The modeled device allows the determination of several of the physical parameters (see table below)

**Table 7.6  Devices physical parameters**

| device | T | $\delta$ | Ro | $C_t$ | $C_d$ | $L_d$ |
|--------|-----|------|-----|------------------|------------------|------------------|
| 1-port | 791 | 0.11 | 138 | ------ | $1 \times 10^{-7}$ | $3.3 \times 10^{-7}$ |
| 2-port #1 | 575 | 0.1 | 350 | $4.2 \times 10^{-8}$ | $1.1 \times 10^{-9}$ | insignificant |
| 2-port#2 | 701 | 0.1 | 50 | $4.2 \times 10^{-8}$ | $1.1 \times 10^{-9}$ | insignificant |

where T is the membrane tension in N/m, $\delta$ is the membrane loss tangent, Ro is the device dc resistance, $C_t$ is the transfer capacitance between the transducers, $C_d$ is the direct capacitance between the device and the substrate and $L_d$ is the inductance.

For the one-port device we have determined to be 71.78 $\Omega$, which is a resistance in parallel with Ro comes from the mismatch between the device impedance and the coaxial cable.

For the experimental flexural plate wave devices, the electrical quality factor Q was determined using two different ways:

a) by taking the ratio of the center frequency and the -3dB frequency bandwidth

b) by the product of the angular resonance frequency and the slope of the phase divided by two.

For the theoretical flexural plate wave devices the mechanical quality factor is given by:

a) $Q = \omega R C$

b) by the product of the angular resonance frequency and the slope of the phase divided by two.

Table 7.7 gives the values of the quality factors for both the experiment and theory..

**Table 7.7   Quality factor for the two-port flexural plate wave device ( device #1)**

| | Experimental | | Theory | |
|---|---|---|---|---|
| | Electro-mechanical (using 3db ) | Electro-mechanical (slope) | Mechanical $Q = \omega R\,C$ | Electro-mechanical (slope) |
| Q | 60 | 56 | 150 | 32 |

Some of the physical parameters such as, membrane bending moment, poisson's ratio, membrane lost tangent, membrane areal mass density, young's moduls and membrane internal stress (tension) for Silicon Carbide and Silicon Nitride films are given by Table 7.8.

**Table 7.8 Silicon Carbide and Silicon Nitride's physical parameters**

| | E (N/m²) x $10^{11}$ | T (N/m) | D (N.m) x $10^{-7}$ | lt | $v$ | $\rho_s$ (kg/m²) x $10^{-2}$ |
|---|---|---|---|---|---|---|
| SiC | 3.92 | 571-790 | 1.1 | 0.1 | 0.14 | 1.27 |
| SiN | 2.76 | 142 | 1.95 | 0.003 | 0.24 | 1.39 |

# CHAPTER VIII

## Summary and Suggestions for Future Research

### 8.1 Summary

Flexural Plate Wave (FPW) devices fabricated from Silicon Carbide (SiC) membranes are presented here which exhibit electrical and mechanical characteristics in its transfer functions that make them very useful as a low voltage probe devices capable of functioning in small areas that are commonly inaccessible to ordinary devices. The low input impedance characteristic of this current driven device makes it possible for it to operate at very low voltages, thereby reducing the hazards for flammable or explosive areas to be probed. The primary objective was to study the suitability of Silicon Carbide (SiC) membrane as a replacement of Silicon Nitride (SiN) membrane in flexural plate wave developed by Sandia National Laboratories. The performance obtained from Silicon Carbide Flexural Plate Wave devices are as expected from the model. The membrane loss tangent was larger than expected due to materials limitations and device fabrication method. It was shown that, for a total reflection the edges of the membrane have to be nearly perfect. We can approach the ideal condition with devices that we have but to get the exact theoretical condition we have to have perfectly fabricated edges and less mechanical stress.

### 8.2  Suggestion For future Research

This work demonstrates the feasibility of using silicon carbide in FPW sensors and resonators. However, in order to fully exploit the potential of using silicon carbide for sensor and resonator applications, additional work could be done, where improvement in the mechanical and geometrical fabricated forms could be made. The application of silicon carbide to acoustic wave devices requires extremely high qualities in the material and extensive clean labs as well as improvements in the fabrication techniques for the FPWs devices. The data obtained here can be used to extrapolate accurately vacuum conditions by letting $\rho_f$ ( density of fluid) approach zero.

# CHAPTER IX

## 9.1 Conclusion

In this work, the fabrication and feasibility of application of a one- and two-port flexural plate wave device have been demonstrated using silicon carbide membranes. The mechanical characteristics of the devices exhibited predictable results, both experimentally and theoretically. A mathematical model was developed to understand the devices' transfer characteristics using, enabling a comparison between the experimental and estimated results. The comparison shows that the one- and two-port devices not only behaved as predicted, but can be flexibly tailored to provide physical interfaces for electrical-mechanical transducers or sensors.

A stress problem had to be overcome to allow the growth of a silicon carbide film with minimum stress. This was accomplished by adjusting the growth parameters. Once that problem was solved, the second problem was the release of a free standing silicon carbide membrane. This problem was overcome by diluting HF as a silicon etch solution, after unsuccessful attempts with KOH and TMAH as etch solutions, and by the use of black wax as an etch mask. Herewith, we were able to release several silicon carbide membrane. The third problem was the alignment of the meander line transducer with respect to the membrane boundaries due to the limited visibility of the chrome-gold layer after evaporation.  This problem was solved after several practice sessions.

The mathematical model enabled us to graphically determine the membrane tension, the membrane loss tangent, the direct and transfer electrical properties of the transducers. We were able to theoretically reproduce our experimental data and also modeled Sandia's one port flexural plate wave device [1],  published in the Journal of Applied Physics paper in 1998 (Figure 9.1).  We obtained a quality factor of 5454.

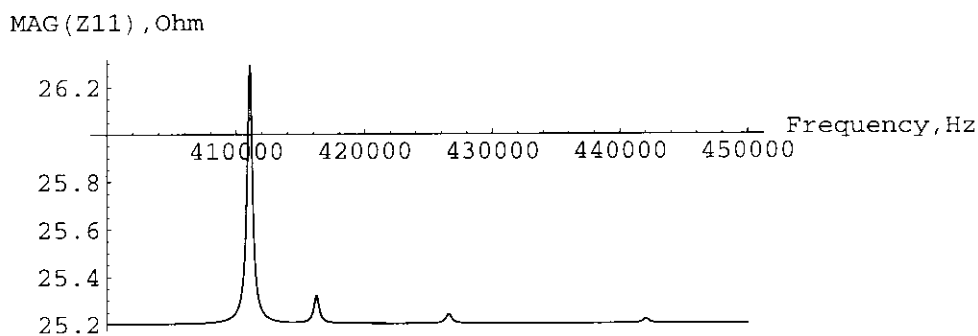

Fig. 9.1 Impedance response of the one-port flexural plate wave device

To the best of our knowledge, there have been no other reports of successful results for any transducers using silicon carbide membranes. We believe that these results will add significantly to any future research in the development of flexural plate wave devices using silicon carbide films.

# ACKNOWLEDGMENTS

NFD would like to dedicate this work to her grandmother Adja Binetou Bineta Fall, her father El Adji Amadou Lamine Diagne, and her mother Mame Caroline Kamara. We are grateful for the kind and professional support that Mr. James Griffin has given me in this research endeavor. A special thanks to Mr. Ebenezer Eshun for his invaluable advice and time during the experimental research. Also, we would like to thank all the members of Material Science Research Center of Excellence. A special thanks to Professor Gary Harris for his support of this research during Professor Spencer's absence.

A special thanks to Dr. Kent Schubert at Sandia National Laboratories and the external committee member (for the dissertation defense)Dr. Steven Martin for their support, time and advice to this research. We am indebted to the Sandia National Laboratories for the financial support they have given to me for this research. We are grateful to the members of the Computational Physics Laboratory for their academic support.

A special thanks to Mr. Edward H. Dowdye for his support, both mentally and spiritually. Also Ms. Helen Major for her support, advice and kindness. A special thanks to Dr. Chandran Haridas for his valuable academic support, and to Ms. Maryann Tann and Ms. Sandra Logan for their support.

# Appendix A

In this section experimental details are provided for each device. Results were obtained using different magnetic field values.

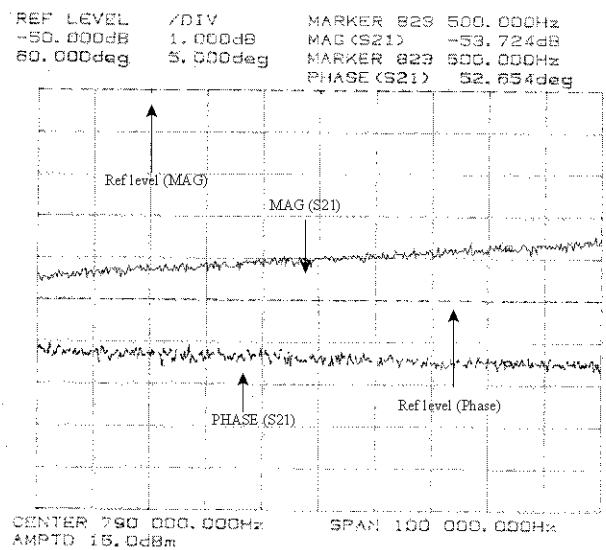

Fig. 7.15  Transmission response (S21)(amplitude and phase) of the Two-port FPW Device (Device **#1**, $B_x = 0$ KG)

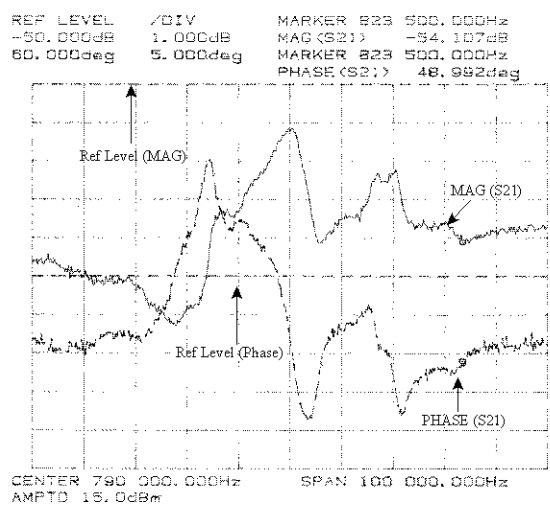

Fig. 7.16 Transmission response S21 (amplitude and phase) of the Two-port FPW Device (Device #1, $B_x$ = **6 KG**)

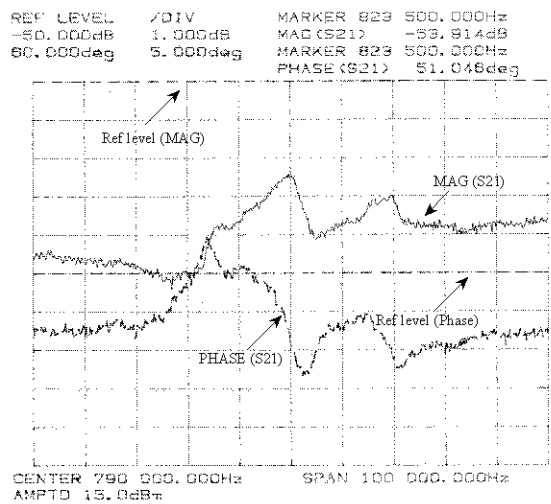

Fig. 7.17    Transmission response (amplitude and phase) of the Two-port FPW Device (Device #1, $B_x$= 8.4 KG)

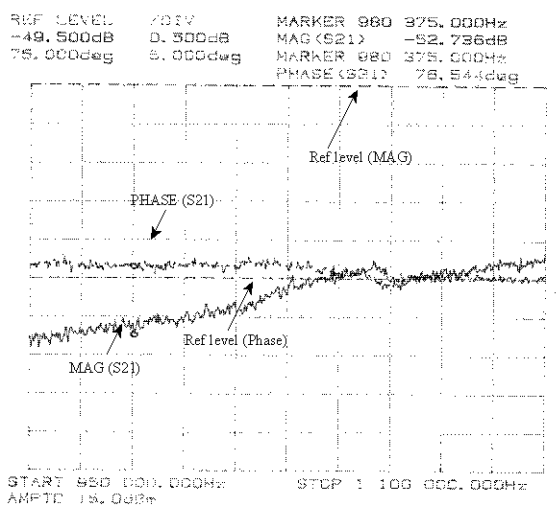

Fig. 7.18    Transmission response S21 (amplitude and
             phase) of the Two-port FPW Device (Device
             **#2**, $B_x$ **= 0 KG)**

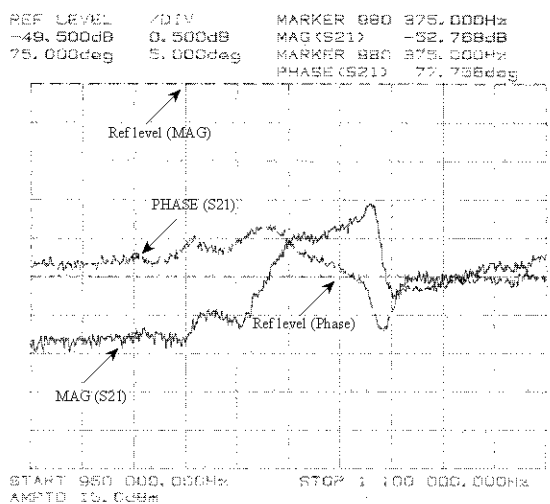

Fig. 7.19    Transmission response S21 (amplitude and
             phase) of the Two-port FPW Device (Device
             #2, $B_x$= 6 KG)

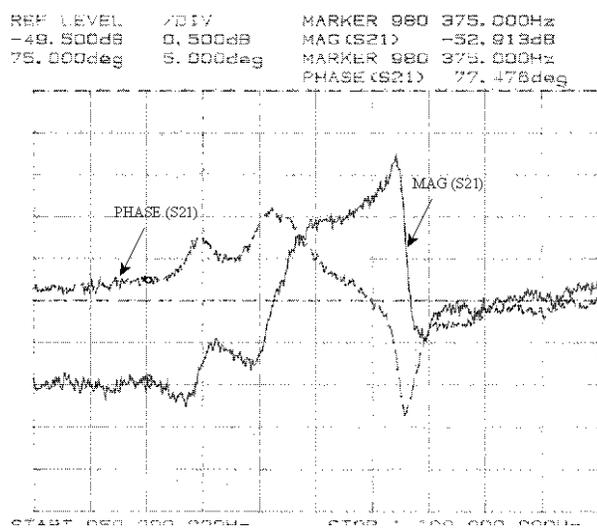

Fig. 7.20    Transmission response (amplitude and phase) of the Two-port FPW Device (Device #2, $B_x$= 8.4 KG)

# Appendix B

Publication

"The development of Resistive Heating for the High Temperature Growth Of Alpha SiC using a vertical CVD Reactor", Ebenezer Eshun[1], Crawford Taylor[1], N. Fama Diagne[1], James Griffin[1], M.G. Spencer[1], Ian Ferguson[2], Alex Gurray[2] and Rick Stall[2], Poster presentation at the ICSCRM '99 Conference, Raleigh NC, October 1999, for publication in the ICSCRM Proceedings.

[1] Howard University, Materials Science Center of Excellence, Washibgton DC 20059

[2] EMCORE Corporation, Somerset, NJ